\documentclass[conference]{IEEEtran}
\usepackage{cite}
\usepackage{amsmath,amssymb,amsfonts}
\usepackage{graphicx}
\usepackage{textcomp}
\usepackage[ruled,linesnumbered,vlined]{algorithm2e}
\usepackage{stfloats}
\usepackage[hyphens]{url}
\usepackage{verbatim}
\usepackage{dsfont}
\usepackage{algpseudocode}
\usepackage{pifont}
\usepackage{color}
\usepackage{soul}
\usepackage{booktabs}
\usepackage{tabularx}
\usepackage{flushend}
\usepackage[normalem]{ulem}
\usepackage[table,xcdraw]{xcolor}
\usepackage{array}
\usepackage{threeparttable}

\usepackage[colorlinks=true,
            urlcolor=black,
            linkcolor=red,
            anchorcolor=black,
            citecolor=green]{hyperref}

\usepackage{makecell}
\usepackage{multirow}
\usepackage{nicematrix}
\usepackage{float}
\usepackage[numbers,sort&compress]{natbib}
\usepackage[justification=justified]{caption}
\usepackage{subfigure}

\begin{document}

\title{SparseLUT: Sparse Connectivity Optimization for Lookup Table-based Deep Neural Networks}

\author{
        {Binglei Lou, Ruilin Wu, Philip Leong}\\
        {School of Electrical and Computer Engineering}\\
       {The University of Sydney, 2006, Australia}\\
       {Email: \{binglei.lou,ruilin.wu,philip.leong\}@sydney.edu.au}    
}

\maketitle

\begin{abstract}
The deployment of deep neural networks (DNNs) on resource-constrained edge devices such as field-programmable gate arrays (FPGAs) requires a careful balance of latency, power, and resource usage while maintaining high accuracy. Existing Lookup Table (LUT)-based DNNs, including LogicNets, PolyLUT, PolyLUT-Add, and NeuraLUT, exploit native FPGA resources with random sparse connectivity. This paper introduces SparseLUT, a connectivity-centric training technique tailored for LUT-based DNNs. SparseLUT leverages a non-greedy training strategy that prioritizes the pruning of less significant connections and strategically regrows alternative ones, resulting in efficient convergence to the target sparsity. Experimental results show consistent accuracy improvements across benchmarks, including up to a 2.13\% increase on MNIST and a 0.94\% improvement for Jet Substructure Classification compared to random sparsity. This is done without any hardware overhead and achieves state-of-the-art results for LUT-based DNNs.
\end{abstract}


\IEEEpeerreviewmaketitle
\section{Introduction}

Deep neural networks (DNNs) have significantly improved our ability to solve pattern recognition problems across diverse data formats, including images, video, speech, audio and text~\cite{lecun2015deep}. Field-Programmable Gate Arrays (FPGAs) provide a unique implementation platform for deploying DNNs as they enable better integration with other system components such as networks and video decoders, allow signal processing to be combined with DNNs, and can operate with lower energy and latency than graphics processing units, particularly in real-time inference tasks.

\begin{figure}[b]
    \centerline{\includegraphics[width=0.4\linewidth]{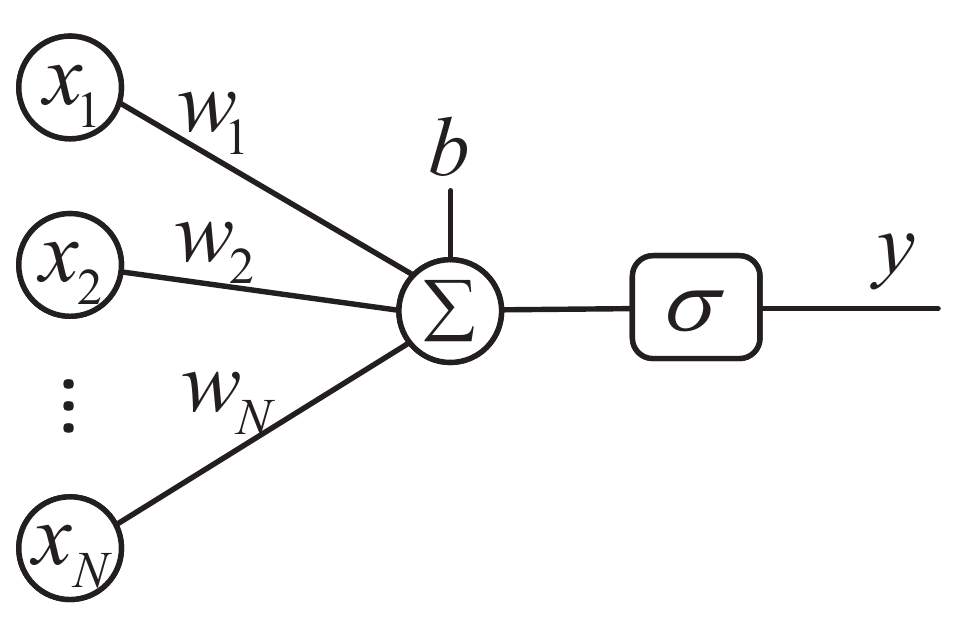}}
    \caption{Neuron computation with fan-in $N$. LUT-DNNs typically select a random subset (size $F \ll N$) of the inputs. SparseLUT is a training scheme to select the inputs while maximizing accuracy.} 
    \label{fg:neuron}
\end{figure}

In an artificial neural network, the output of a neuron is given by 
\begin{equation}
y = \sigma \left( {\sum\nolimits_{k = 1}^{N} {{w_k}{x_k} + b} } \right)
\label{eq:neuron}
\end{equation}
where $N$ is the number of inputs, $x_i$ are the inputs, $w_i$ are the weights, $b$ is a bias term and $\sigma(\cdot)$ is the activation function. 
Lookup Table (LUT)-based DNNs (LUT-DNN) exploit FPGA-native elements to achieve exceptional area efficiency and latency. This is done by combining the multiplication, sum, and activation operations of Figure~\ref{fg:neuron} (Equation~\ref{eq:neuron} in a single LUT). Recent approaches, including LogicNets~\cite{LogicNets}, PolyLUT~\cite{polylut}, PolyLUT-Add~\cite{polylutadd}, and NeuraLUT~\cite{neuralut}, introduce \textit{a priori} weight sparsity by constraining each neuron's fan-in, $F$. Put another way, we constrain the weight vector $W = \{{w_1},{w_2}, \ldots ,{w_N}\} \in {\mathbb{R}^{{N}}}\label{eq:W}$ to be determined \textit{a priori}, sparse and have a maximum of $F$ non-zero values ($0 < F \ll N$). To note, the terms sparsity and connectivity are used interchangeably in this paper.

Weight sparsity is crucial for managing resource consumption in LUT-DNNs as the hardware cost grows exponentially with $F$. 
Previous works choose the $F$ out of $N$ connections randomly~\cite{LogicNets,polylut,polylutadd,neuralut}, inherently limiting the potential of such networks to achieve higher accuracy. 
While some methods, such as post-training pruning~\cite{lecun1989optimal,thimm1995evaluating,gale2019state,sze2020efficient} and sparsity optimization during training~\cite{gmp,deepr,rigl,srigl}, introduce sparsity and have proven effective for conventional DNNs, they control sparsity at the model or layer level and do not address the unique challenges of LUT-based neural networks with neural-level fan-in constraints. This highlights a significant gap and opportunity. 

\begin{figure*}[h]
    \centerline{\includegraphics[width=0.8\linewidth]{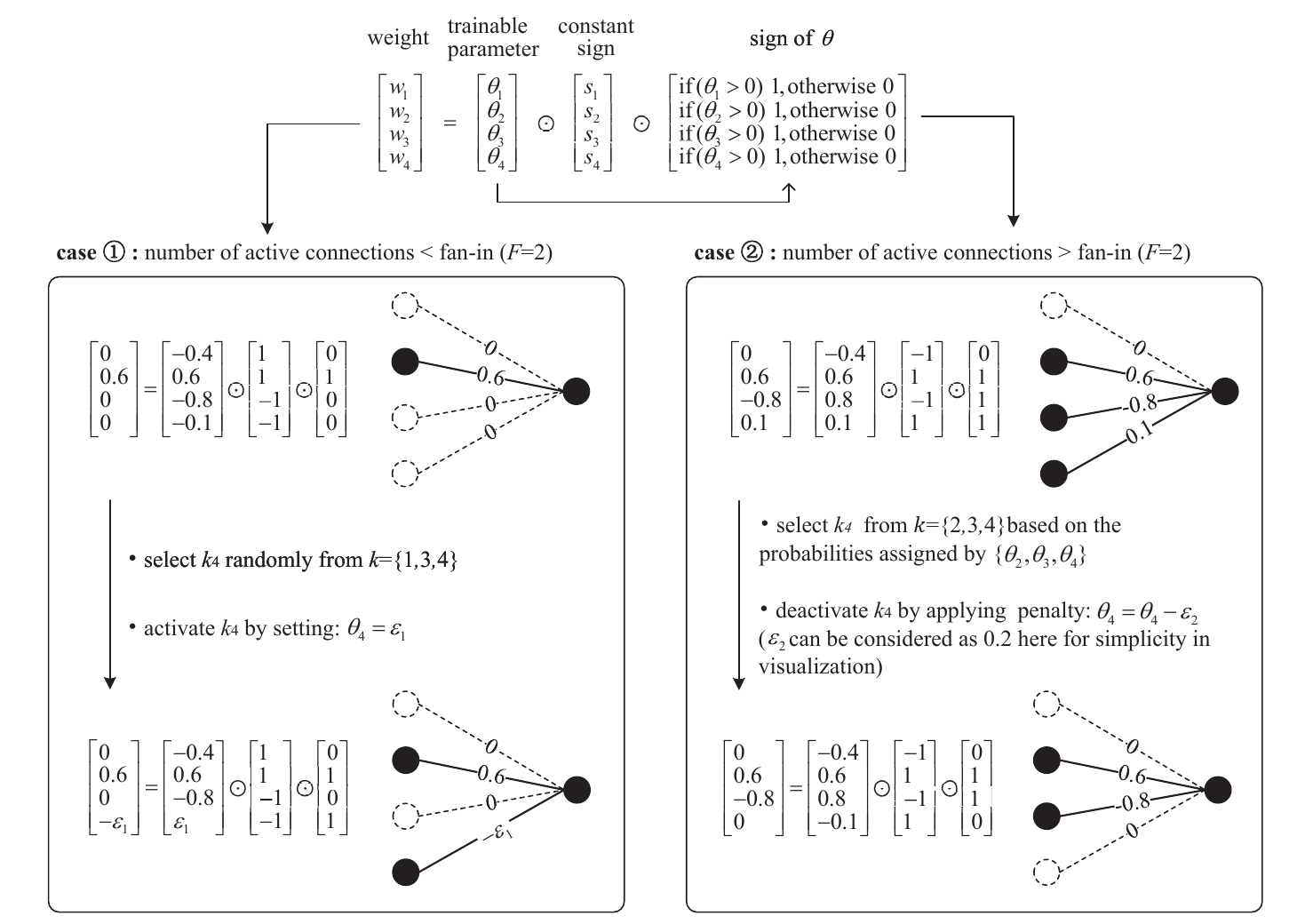}}
    \caption{Example iteration showing SparseLUT adjusting neuron connections to reach a target fan-in of 2, introducing new ones to increase the fan-in from 1 to 2 and eliminating connections to reduce the fan-in from 3 to 2. Note the \( \epsilon_2 \) parameter is typically set to a much smaller value, requiring multiple iterations to deactivate a connection.} 
    \label{fg:example_f2}
\end{figure*}

SparseLUT addresses the unique challenges outlined through careful selection of the non-zero $w_k$ values in Equation~\ref{eq:neuron}. Figure~\ref{fg:example_f2}, to be elaborated in Section~\ref{se:Methodology}, illustrates our approach to training wherein active connections are gradually modified to ultimately achieve a target fan-in of $F=2$. As training progresses, new connections are added to neurons with fewer than \( F \) active connections (Case~\ding{172}), while neurons with active connections exceeding \( F \) (Case~\ding{173}) are deactivated. Over the training process, the target fan-in is achieved for all neurons.

The main contributions of this work are as follows: 

\begin{enumerate}
\item We introduce the first architecture for LUT-based neural networks that optimizes connectivity of neuron inputs during training. It can be applied to any LUT-DNN without LUT or routing resource overhead as it only affects how the $F$ inputs in Figure~\ref{fg:neuron} are chosen.

\item We present SparseLUT, an accompanying non-greedy training algorithm for sparsifying dense or sparse networks. This is a modification of the Deep Rewiring (DeepR)~\cite{deepr} approach where we ensure each neuron has $F$ inputs and relax the requirement that the number of dropped and regrown connections must always match. 

\item We evaluate our method on a diverse set of representative LUT-DNN models, including LogicNets~\cite{LogicNets}, PolyLUT~\cite{polylut}, PolyLUT-Add~\cite{polylutadd}, and NeuraLUT~\cite{neuralut}. Compared to the original designs, our optimized connectivity consistently achieves superior accuracy across all evaluated baselines. The work in this paper is reproducible to facilitate experimentation with our design. Source code and data to reproduce our results are available from Github\footnote{SparseLUT: \url{https://github.com/bingleilou/SparseLUT}}. 
\end{enumerate}

The remainder of this paper is organized as follows: Section~\ref{se:Background} reviews related work, including LUT-based DNNs and deep sparsification techniques. Section~\ref{se:Methodology} describes the proposed method in detail, including the neuron-level sparsity optimization and non-greedy sparsification strategies. Section~\ref{se:Results} presents experimental results, showing accuracy and resource utilization improvements over existing methods. Finally, Section~\ref{se:Conclusion} concludes the paper and outlines future research directions.

\section{Background}
\label{se:Background}
\subsection{\textit{A priori} Random Sparsity DNN for LUT inference}

\begin{figure*}
    \centerline{\includegraphics[width=0.82\linewidth]{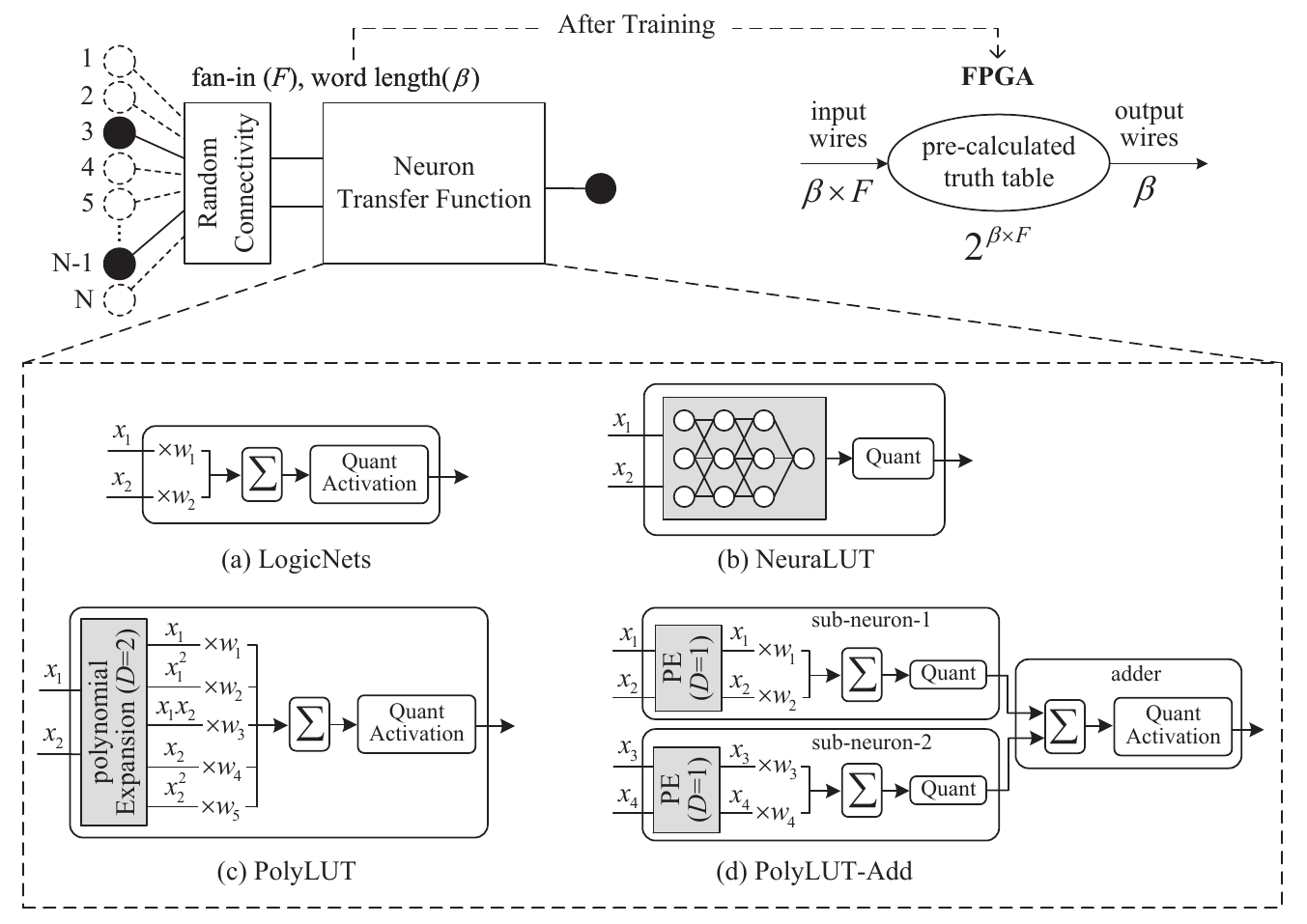}}
    \caption{Illustration of the generalized LUT-DNN architectures with sparse connectivity.} 
    \label{fg:lut}
\end{figure*}

The top part of Figure~\ref{fg:lut} shows a generalized LUT-DNN approach; a maximum of $F$ inputs are randomly selected from ${N}$ nodes of the current layer to connect to each neuron of the next layer (only a single neuron of the next layer is demonstrated here). Additionally, the bit width of each neuron's inputs and outputs are quantized as $\beta$, and the rest of the parameters inside the neuron transfer function are maintained in full precision. Therefore, after training, the transfer function mapping an input vector $\left[ {{x_1},{x_2}, \ldots ,{x_{F}}} \right]$ to the output node can be implemented using $\beta F$ inputs in hardware, and hence its implementation requires $\mathcal{O}(2^{\beta F})$ LUTs in a pre-calculated truth table. 

Several LUT-DNN approaches have been proposed and are summarised in the bottom part of Figure~\ref{fg:lut}. LogicNets~\cite{LogicNets}, introduced by Umuroglu \emph{et al.}, incorporates a sum of products and activation functions in the neuron transfer function. Techniques to improve the accuracy of LogicNets can be split into two categories: (1) improve the representation ability of each LUT, and (2) improve the fan-in of each neuron. PolyLUT~\cite{polylut} and NeuraLUT~\cite{neuralut} proposed by Andronic~\emph{et al.} utilize the former approach. Specifically, PolyLUT enhanced accuracy by introducing piecewise polynomial functions. NeuraLUT~\cite{neuralut} employs the network-in-network technique~\cite{nin} and, therefore, maps entire sub-nets to a single LUT, enabling a deeper NN and better accuracy. As the polynomial computations and sub-nets are absorbed within the LUT, the number of entries in the lookup tables generated does not change. In contrast, the PolyLUT-Add~\cite{polylutadd} proposed by Lou~\emph{et al.} can be categorized as the latter approach, addressing the fan-in constraint of LUT-based DNNs using multiple PolyLUT neurons. The technique enhances neuron connectivity by combining $A$ PolyLUT sub-neurons via addition to improve accuracy ($A=2$ and polynomial expansion (PE) with $D=1$ is used in this example). It generates multiple decoupled truth tables to reduce the total number of LUTs to $\mathcal{O}(A \times 2^{\beta F} + 2^{A(\beta +1)})$ with its later $\mathcal{O}(2^{A(\beta +1)})$ LUTs consumed by an adder layer.

\subsection{Sparsification}

\begin{table*}[b]
\centering
  \caption{Summary of dynamic sparsification approaches.}
  \def\arraystretch{1.15}
  %\scalebox{1.0}{
  \setlength{\tabcolsep}{0.3cm}
  \resizebox{0.98\linewidth}{!}{ 
  \begin{tabular}{|c|c|c|c|c|}
  %\toprule
  \hline
  \multicolumn{1}{|c|}{\textbf{Method}}    & \multicolumn{1}{c|}{\textbf{Training}} & \multicolumn{1}{c|}{\textbf{Drop (Prune) $\searrow$}} & \multicolumn{1}{c|}{\textbf{Regrowth $\nearrow$}} & \multicolumn{1}{c|}{\textbf{\textit{a priori} fixed fan-in constrain}}\\ \hline \hline
  GMP~\cite{gmp}       & dense-to-sparse   & magnitudes  & \ding{55}    & \ding{55}   \\
  \hline
  DeepR~\cite{deepr}   & sparse-to-sparse  & stochastic  & random      & \ding{55}   \\
  \hline
  RigL~\cite{rigl}     & sparse-to-sparse  & magnitudes  & gradient    & \ding{55}   \\
  \hline
  SRigL~\cite{srigl}   & sparse-to-sparse  & magnitudes  & gradient    & \ding{55}   \\
  \hline
  SparseLUT (ours)    
  & flexible          & stochastic+magnitudes  & random  & \ding{51}   \\
  \hline
\end{tabular}}
\label{tb:dst_review}
\end{table*}

In practice for most dense DNNs, a significant fraction of weights are redundant and removable~\cite{sze2020efficient}. Research on sparsifying neural networks, known as network pruning, dates back to the late 1980s and mid-1990s. These techniques begin with a pre-trained, densely connected model and perform weight removal, where unimportant weights are identified and set to zero. Various approaches have been used as the removal criteria. For example, early work in 1989, Optimal Brain Damage~\cite{lecun1989optimal}, removed non-salient weights as determined via the diagonals of the Hessian matrix. In 1995, Thimm \emph{et al.}~\cite{thimm1995evaluating} demonstrated that removing weights based on their magnitude was a simple yet effective technique. Gale \emph{et al.}~\cite{gale2019state} further examined alternative pruning criteria on large-scale learning of transformers trained on WMT2014 English-to-German and ResNet-50 trained on ImageNet. They concluded that complex techniques such as $L_0$ regularization~\cite{louizos2017learning} and Variational Dropout~\cite{molchanov2017variational} achieve similar accuracy-sparsity trade-offs compared to magnitude-based pruning. 

Gale \emph{et al.}~\cite{gale2019state} also observed that unstructured sparse architectures learned through pruning alone cannot achieve as good performance as a model trained with joint sparsification and optimization. Thus
more recent methods commence from scratch rather than a pre-trained model, \emph{i.e.}, they (1) start from the original initial training conditions and (2) update connectivity by doing weight removal and weight re-growth (optional) simultaneously.
This can be further divided into two subcategories with representatives listed in Table~\ref{tb:dst_review}: (1) The training starts with a dense model and gradually removes connections during training to achieve the desired sparsity~\cite{gmp}. (2) The training begins with a sparse model to minimize training memory overhead, and during training, connections are dropped and regrown during training, guided by criteria such as weight magnitudes, gradients, or stochastically~\cite{rigl,srigl,deepr}.

Gradual Magnitude Pruning (GMP)~\cite{gmp}, removes connectivity from a dense initial solution during training until the desired sparsity is reached. It begins with a dense network and removes weights with the smallest magnitudes. Its sparsity decreases monotonically until the target sparsity is reached. 

In contrast, there are approaches that employ dynamic rewiring of neural networks~\cite{deepr, rigl, srigl}. This strategy is inspired by the human brain, where the majority of brain volume is occupied by white matter—the network of connections between neurons. In the brain, synaptic connectivity is highly dynamic, with the continuous formation and reorganization of synapses, particularly during learning~\cite{holtmaat2005transient, stettler2006axons, attardo2015impermanence, chambers2017stable}.

Deep Rewiring (DeepR)~\cite{deepr} introduces a stochastic framework for dynamic rewiring. At initialization, DeepR employs a network with a limited number of random connections. During training, at each iteration, connections are removed if their weight updates lead to a sign change. A new connection is then randomly activated to preserve the target sparsity level. Our work extends this random walk mechanism, and further details are provided in subsequent sections.
Rigged Lottery (RigL)~\cite{rigl} removes weights with the smallest magnitudes and regrows connections based on gradient magnitude during training. At each iteration, the number of dropped connections equals the number of regrown connections, maintaining a consistent sparsity level. RigL is designed to handle standard unstructured sparsity constraints at the layer level.
Structured RigL (SRigL)~\cite{srigl} extends RigL by incorporating constant fan-in structured sparsity, which facilitates a compact weight matrix representation. Specifically, after training, the non-zero weights in each matrix exhibit a rectangular pattern. SRigL achieves its target fan-in target on average, rather than this being enforced on a per-neuron basis.

\section{Methodology}
\label{se:Methodology}
\subsection{Model Structure}

Unlike conventional DNNs, which focus solely on updating the network weights and biases, SparseLUT also incorporates weight removal and regrowth under a set of predefined constraints. A key factor enabling this is its weight representation. We follow the approach of DeepR~\cite{deepr} where each connection \( k \) in the network is represented by two components:

\begin{itemize}
    \item A \textbf{trainable connection parameter} $\theta_k \in \mathbb{R}$, which controls both the magnitude of the connection and its status (active or inactive).
    \item A \textbf{predefined, non-trainable constant sign} $s_k \in \{-1, 1\}$, initialized randomly, which encodes the fixed polarity of the connection.
\end{itemize}

Referring to Figure~\ref{fg:example_f2}, we use $W = \{{w_1},{w_2}, \ldots ,{w_N}\} \in {\mathbb{R}^{{N}}}$ to denote the weight vector of a neuron in Equations~\ref{eq:neuron}.

Algorithm~\ref{alg:sparselutweight} outlines the procedure for determining the effective weight vector for each neuron in SparseLUT. The algorithm begins with an initial weight vector \( W_0 \in \mathbb{R}^N \), where the magnitude of each element is used as the initial value for \(\theta_k\). This ensures dense connectivity under the initial conditions, with all connections corresponding to \(\theta_k\) being non-negative. The initial sparsity of the network is determined by the element-wise product of \(|W_0|\) and a binary sparsity mask vector \(\textit{is\_con}\), where only the element $F_i$ is set to 1, and all other elements are set to 0. To finalize the effective weight vector \( W \), an indicator function \(\text{1}(\cdot)\) is applied to the product \(\theta_k \odot s_k\), controlling the active status of each connection.

\begin{algorithm}
\caption{Weight Mapping Procedure}
\label{alg:sparselutweight}
\KwIn{Weight dimensions for $W$, initial fan-in $F_i$.}
\BlankLine
$W_0 \gets$ Randomly initialize weights $W_0$ from a standard normal distribution with the same dimensions as $W$\;
$is\_con \gets$ Generate a binary sparsity mask $is\_con$ such that each output neuron is connected to $F_i$ randomly chosen input neurons\;
$\theta \gets |W_0| \odot is\_con$ {(Compute the initial $\theta$ values as the element-wise product)}\;
$sign \in \{-1, +1\}$ Randomly initialize weight signs with uniform probability\;
$W \gets \theta \cdot sign \cdot \text{1}(\theta > 0)$ {(Compute the final weights, where $\text{1}(\cdot)$ is the indicator function that sets negative values of $\theta$ to zero)}\;
\BlankLine
\KwRet{Final weight vector $W$}
\end{algorithm}

\subsection{Model Training}

\begin{algorithm}
\caption{SparseLUT Training Procedure}
\label{alg:sparselut}
\KwIn{Target fan-in $F_o$, sparse-to-sparse training starting point $T$, learning rate $\eta$, regularization coefficient $\alpha$, and noise ${v_k} \sim N(0,G^2)$.}
\KwOut{Feature mask $\mathcal{M}$}
\BlankLine
\textbf{Initialization:}\\
$W \gets $ defined in Algorothm~\ref{alg:sparselutweight}\;
\BlankLine
\textbf{Training:}\\
\For{each training step $t$}{
    
    \For{all active connections $k$ ($\theta_k>0$)}{
        Update $\theta_k \leftarrow \theta_k - \eta \frac{\partial E(\theta)}{\partial \theta_k} - \eta \alpha + \eta {v_k}$\;
        \textbf{if} $\theta_k < 0$ \textbf{then} set connection $k$ non-active \;
    }
    % Adjust connections to meet target fan-in
    $R \leftarrow$ number of active connections - $F_o$\;
    \eIf{$R<0$}{
        Select $\left| R \right|$ non-active connections $k'$ with uniform probability and activate them\;
        Set $\theta_{k'} \leftarrow \epsilon_{1}$\;
    }
    {
        \eIf{$t<T$}{
            Rank the \( \theta \) values of all active connections\;  
            Select \( \left| R \right| \) active connections \( k' \) with the lowest rank probabilities and apply penalties\;
            Update ${\theta _{k'}} \leftarrow {\theta _{k'}} - \epsilon_{2}$\;
        }
        {
            Rank the \( \theta \) values of all active connections\;  
            Select $\left| R \right|$ active connections $k'$ with the lowest rank probabilities and deactivate them\;
            Set ${\theta _{k'}} \leftarrow 0$\;
        }
    }
}
\BlankLine
\KwRet Feature mask $\mathcal{M}$, where $\mathcal{M}_k = 1$ if $\theta_k > 0$, otherwise $\mathcal{M}_k = 0$\;
\end{algorithm}

The SparseLUT training procedure, as outlined in Algorithm~\ref{alg:sparselut} is a non-greedy algorithm. Each iteration updates the connection parameters and adjusts the network's sparsity to enforce the target fan-in \( F_o \) sparsity during training.

At each backpropagation training step~\cite{Sun_2019_sgdsurvey}, only the parameters \( \theta_k \) of active connections (\( \theta_k > 0 \)) are updated according to the rule in Line~6. Given that the learning rate is $\eta$, the term \( \eta \frac{\partial E(\theta)}{\partial \theta_k} \) represents the derivative of the error function, while \( \eta \alpha \) is regularization term ($\alpha$ is any regularization function). The final term, \( \eta {v_k} \), originally introduced in DeepR~\cite{deepr}, implements a random walk scheme in the parameter space. Here, \( {v_k} \sim N(0, G^2) \) denotes a noise matrix with the same dimensions as \( \theta_k \), a mean of \( 0 \), and a standard deviation of \( G \). Following this update, any connection for which \( \theta_k \) becomes negative is deactivated.

For each neuron, the difference between the existing number of active connections and the target fan-in is computed as \( R \). A negative \( R \) indicates that the number of active connections is smaller than the target fan-in, prompting the activation of \( |R| \) inactive connections \( k' \) selected uniformly at random (Line~10). These connections are initialized with a small value \( \epsilon_1 \) to allow them to rejoin the backpropagation process without dominating the updates immediately after reactivation.

Conversely, a positive \( R \) means that \( |R| \) active connections need to be deactivated to meet the fan-in constraint. Different from approaches in Table~\ref{tb:dst_review}, SparseLUT employs a two-phase training strategy for this deactivation process:  

\begin{enumerate}
    \item {progressive sparsification phase:}  
    During the early stages of training (\( t < T \)), a small penalty term \( \epsilon_2 \) is subtracted from \( \theta_{k'} \) in each iteration (Line~16). This gradual penalty reduces the values of smaller parameters to a greater extent, under the assumption that connections with smaller weights contribute less to the neuron’s output. As \( \theta_{k'} \) becomes negative, these connections are eliminated. This stage relaxes the strict requirement that the number of dropped and regrown connections must always match, as used in conventional regrowth approaches such as DeepR~\cite{deepr}, RigL~\cite{rigl}, and SRigL~\cite{srigl}. We will later show that this greater flexibility during training leads to consistently improved accuracy.

    \item {fine-tuning phase:}  
    In the later stage of training (\( t \geq T \)), the model strictly enforces the fan-in constraint. Connections are directly deactivated when \( |R| > 0 \) (Line~20). This phase focuses on fine-tuning the remaining connections through periodically rewiring eliminated connections and selecting the same number of alternatives to maintain the fan-in constraint for each neuron.  
\end{enumerate}

Upon completing the training process, the final generated output of the algorithm is formed as the feature mask \( \mathcal{M} \), where each element \( \mathcal{M}_k \) indicates the status of connection \( k \) (Line~21).

\subsection{SparseLUT Training Flow and Dynamics}

\begin{figure}[h]
    \centerline{\includegraphics[width=0.50\linewidth]{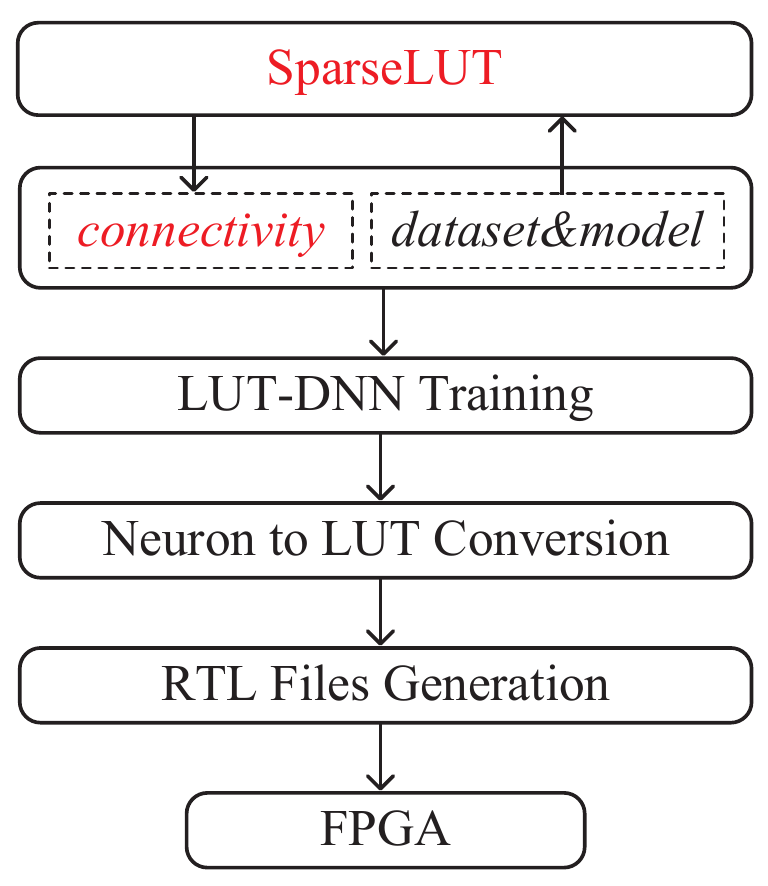}}
    \caption{Workflow of SparseLUT.} 
    \label{fg:toolflow}
\end{figure}

Figure~\ref{fg:toolflow} illustrates the overall workflow of the SparseLUT framework, which can be easily applied to existing LUT-DNN designs~\cite{LogicNets,polylut,polylutadd,neuralut}. SparseLUT begins with the same DNN model configuration, ensuring compatibility with the LUT-DNN training pipeline. It generates the connectivity feature mask \( \mathcal{M} \), compatible with existing LUT-DNNs, and directly replaces the random connectivity in Figure~\ref{fg:example_f2}. The subsequent steps, including DNN re-training, neuron-to-LUT conversion, and RTL file generation, remain unchanged, facilitating smooth integration and ensuring there is no additional hardware overhead.

\section{Results}
\label{se:Results}

\subsection{Datasets}
We evaluated the proposed SparseLUT design using two widely adopted datasets for ultra-low latency inference. These datasets align with the evaluation benchmarks of existing techniques, including PolyLUT~\cite{polylut}, PolyLUT-Add~\cite{polylutadd}, and NeuraLUT~\cite{neuralut}.

\begin{enumerate}
\item \textit{Handwritten Digit Recognition}: In timing-critical sectors such as healthcare, medical imaging, and real-time object tracking, low-latency is crucial. These applications underscore the necessity for swift and accurate decision-making, where even minimal delays can have significant repercussions. Unfortunately, there is no public dataset specialized for low-latency image classification tasks so the MNIST~\cite{mnist} is utilized to compare our work with other LUT-based DNNs. MNIST is a dataset for handwritten digit recognition tasks with $28 \times 28$ pixels as input and 10 classes as outputs.

\item \textit{Jet Substructure Classification}:  Real-time decision-making is often important for physics experiments such as the CERN Large Hadron Collider (LHC). Jet Substructure Classification (JSC) is one of its applications that requires high-throughput data processing. Prior works~\cite{ngadiuba2020compressing,duarte2018fast,coelho2021automatic,fahim2021hls4ml} employed neural networks on FPGA for this task to provide real-time inference capabilities. We also use the JSC dataset formulated from Ref.~\cite{duarte2018fast} to evaluate our work, with the dataset having 16 substructure properties as input and 5 types of jets as outputs. FPGA-based classification in this task must be pipelined to manage a data rate of 40 MHz while ensuring the response latency remains under microsecond.

\end{enumerate}

\subsection{Experimental Methodology}

To evaluate the proposed SparseLUT as a general approach for improving the accuracy of LUT-based DNNs, we tested its performance across several state-of-the-art methods, namely LogicNets~\cite{LogicNets}, PolyLUT~\cite{polylut}, PolyLUT-Add~\cite{polylutadd}, and NeuraLUT~\cite{neuralut}. In particular we compare the effectiveness of random connectivity versus the optimized connectivity of SparseLUT.

The experiment comprises two steps: (1) derive the optimized connectivity feature mask \( \mathcal{M} \). (2) replace the original random connectivity in the baseline models with \( \mathcal{M} \) and retrain the baselines.

\begin{table*}[]
  \centering
  \caption{Setups of baseline models.}
  \label{tb:baselinesetup}
  \bgroup
  \def\arraystretch{1.16}
  \setlength{\tabcolsep}{0.49cm}
  \resizebox{0.99\linewidth}{!}{ 
  \begin{tabular}{|c|l|lc|lcc|}
  \cline{1-7}
  \multicolumn{1}{|c|}{\textbf{Dataset}}    & \multicolumn{1}{c|}{\textbf{Method}} & \multicolumn{1}{c}{\textbf{Model}} & \multicolumn{1}{c|}{\textbf{Neurons per layer}} & \multicolumn{1}{c}{\textbf{$\beta$}} & \multicolumn{1}{c}{\textbf{$F$}} & \multicolumn{1}{c|}{\textbf{$D$}}\\ \hline \hline
  \multirow{3}{*}{MNIST}                 & PolyLUT      & HDR             & 256, 100, 100, 100, 100, 10 & 2       &  6  & 1,2  \\
                                         & PolyLUT-Add  & HDR-Add2        & 256, 100, 100, 100, 100, 10 & 2       &  4  & 1,2  \\
                                         & NeuraLUT     & HDR-5L          & 256, 100, 100, 100, 10      & 2       &  6  & ---  \\\hline\hline
  \multirow{3}{*}{\shortstack{JSC \\ (high accuracy)}}   & PolyLUT      & JSC-XL$^1$ & 128, 64, 64, 64, 5          & 5       &  3  & 1,2     \\
                                         & PolyLUT-Add  & JSC-XL-Add2$^2$   & 128, 64, 64, 64, 5          & 5       &  2  & 1,2  \\
                                         & NeuraLUT     & JSC-5L$^1$ & 128, 128, 128, 64, 5        & 4       &  3  & ---  \\ \hline\hline
   \multirow{3}{*}{\shortstack{JSC \\ (low accuracy)}}   & PolyLUT      & JSC-M Lite      & 64, 32, 5                   & 3       &  4  & 1,2   \\
                                         & PolyLUT-Add  & JSC-M Lite-Add2 & 64, 32, 5                   & 3       &  2  & 1,2      \\
                                         & NeuraLUT     & JSC-2L          & 32, 5                       & 4       &  3  & ---   \\ \hline

  \end{tabular}}
  \egroup

  \begin{tablenotes}
    \footnotesize
    \item[1] PolyLUT with $D=1$ is equivenlent to LogicNets.
    \item[2] Remarks:  $^1$ $\beta_{i}=7$, $F_{i}$ = 2;  $^2$ $\beta_{i}=7$, $F_{i}$ = 1; 
\end{tablenotes}
\end{table*}

To ensure a fair comparison, the configurations and training hyperparameters in Step 2 are consistent with the baseline setups as reported in the original papers and summarized in Table~\ref{tb:baselinesetup}. Note that polynomial degree \( D=1 \) and \( D=2 \) correspond to the linear and quadratic representations, respectively in PolyLUT. LogicNets is a special case of PolyLUT where \( D=1 \), and we utilize \( D \in \{1,2\} \) to evaluate the performance of LogicNets, PolyLUT, and PolyLUT-Add. 

The hardware evaluation is compiled using Vivado 2020.1 on the \texttt{xcvu9p-flgb2104-2-i} FPGA part; the \texttt{Flow\_PerfOptimized\_high} settings configured to perform synthesis in the \texttt{Out-of-Context} (\texttt{OOC}) mode and Default values were used as the Place \& Route settings. All results in this section are obtained by retraining the models using their publicly available implementations.

\begin{table}[]
\centering
  \caption{Sparsity Modes}
  \scalebox{1.0}{
  \renewcommand{\arraystretch}{1.105}
  \setlength{\tabcolsep}{0.6cm}
  \resizebox{0.95\columnwidth}{!}{ 
  \begin{tabular}{c|c}
  \hline
  \multicolumn{1}{|c|}{\textbf{Mode}}    & \multicolumn{1}{c|}{\textbf{Fixed fan-in constrain}} \\ \hline \hline
  \multicolumn{1}{|c|}{Fully Connected}  &  \multicolumn{1}{c|}{---}     \\ \hline
  \multicolumn{1}{|c|}{Random Sparsity}  &  \multicolumn{1}{c|}{\ding{51}}     \\ \hline
  \multicolumn{1}{|c|}{DeepR$^*$~\cite{deepr}}  &  \multicolumn{1}{c|}{\ding{51}}     \\ \hline
  \multicolumn{1}{|c|}{SparseLUT (ours)}  &  \multicolumn{1}{c|}{\ding{51}}     \\ \hline
\end{tabular}}}
\label{tb:4mode}
\end{table}

In Step 1, we evaluate three sparse modes with a fully connected baseline, as shown in Table~\ref{tb:4mode}:  
\begin{itemize}
    \item \textbf{Random Sparsity}: Reflects the approach currently employed by existing techniques~\cite{LogicNets,polylut,polylutadd,neuralut}.  
    \item \textbf{DeepR$^*$}: The original DeepR~\cite{deepr} cannot be directly applied to our problem. We have developed a revised version DeepR$^*$ that supports fixed fan-in, which we use as a comparison baseline.
    \item \textbf{SparseLUT}: Reflects the proposed method.
     
\end{itemize}

All modes follow the same training setup to ensure consistency. The AdamW optimizer~\cite{PyTorch} in PyTorch is used, and each model is trained for 300 epochs. The threshold $T$ is set to 240 epochs, corresponding to 80\% of the total training duration. The hyperparameters $\epsilon_{1} = 10^{-12}$ and $10^{-5} \leq \epsilon_{2} \leq 10^{-4}$ are employed. The $F_i$ is set to $N$ to set the training starts from a dense model.
The primary objective is to obtain the optimized sparsity pattern $\mathcal{M}$. Full-precision training is used for faster convergence and better precision in identifying critical connections. The results presented in this section are based on these configurations, though further optimization or additional training techniques may enhance performance for specific applications.

\subsection{Case Study: MNIST}

We conducted a case study on the MNIST dataset, focusing on two analyses: (1) visualizing the connectivity learned by SparseLUT using heatmaps to evaluate its quality and (2) comparing test accuracy between LUT-DNN models utilizing learned versus random connectivity.

\subsubsection{Connectivity Distribution Visualization}

We lack ground truth to evaluate connectivity quality before testing its performance in LUT-DNNs. However, for the MNIST dataset, where handwritten digits typically appear centered~\footnote{The handwritten digits dataset visualization: \raisebox{-0.25\height}{\includegraphics[height=1.8em]{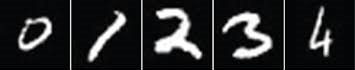}}}, it is reasonable to expect the most important connections to concentrate in the central region.

We adopted a fully connected model and three sparse models (random, DeepR$^*$, SparseLUT) with layer neurons \{256, 100, 100, 100, 10\}, consistent with the HDR-5L model from Table~\ref{tb:baselinesetup}. After training, we averaged the absolute values of the first-layer weight matrix $|W| \in \mathbb{R}^{784 \times 256}$ along the second axis and reshaped it into $W_{ave} \in \mathbb{R}^{28 \times 28}$, aligning with the input image dimensions.

\begin{figure*}[]
    \centerline{\includegraphics[width=0.86\linewidth]{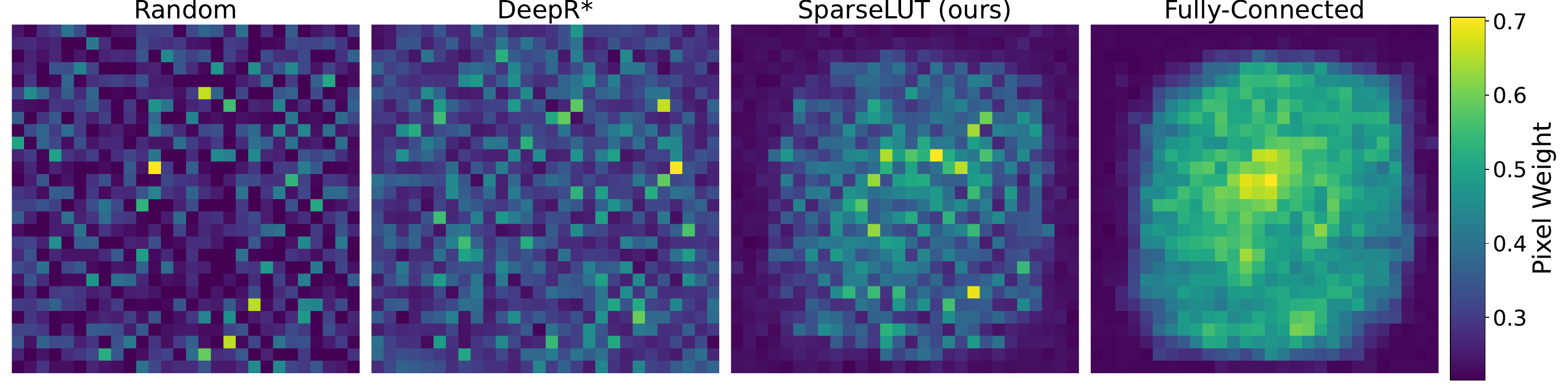}}
    \caption{Heatmaps of the average weight matrix for the first layer in three sparse modes: Random Sparsity, DeepR$^*$, and SparseLUT, with a Fully Connected mode as a baseline. } 
    \label{fg:mnist_heatmaps}
\end{figure*}

Figure~\ref{fg:mnist_heatmaps} is a visualization of $W_{ave}$ as heatmaps. The fully connected model (last sub-figure) shows higher weights concentrated in the central region, aligning with our assumptions.

The first three sub-figures in Figure~\ref{fg:mnist_heatmaps} show heatmaps for sparse models, where only $F=6$ elements in each row of $|W|$ are non-zero. The random sparsity model shows a uniform distribution, reflecting unstructured connectivity. DeepR$^*$ exhibits a trend of concentrated central connectivity, suggesting meaningful adaptation to the task. SparseLUT’s heatmap more closely resembles the fully connected case, indicating that SparseLUT effectively learns an optimized connectivity pattern, prioritizing the central region in alignment with the dataset's structure.

\subsubsection{Test Accuracy Analysis}

\begin{figure}[t]
    \centerline{\includegraphics[width=1.0\linewidth]{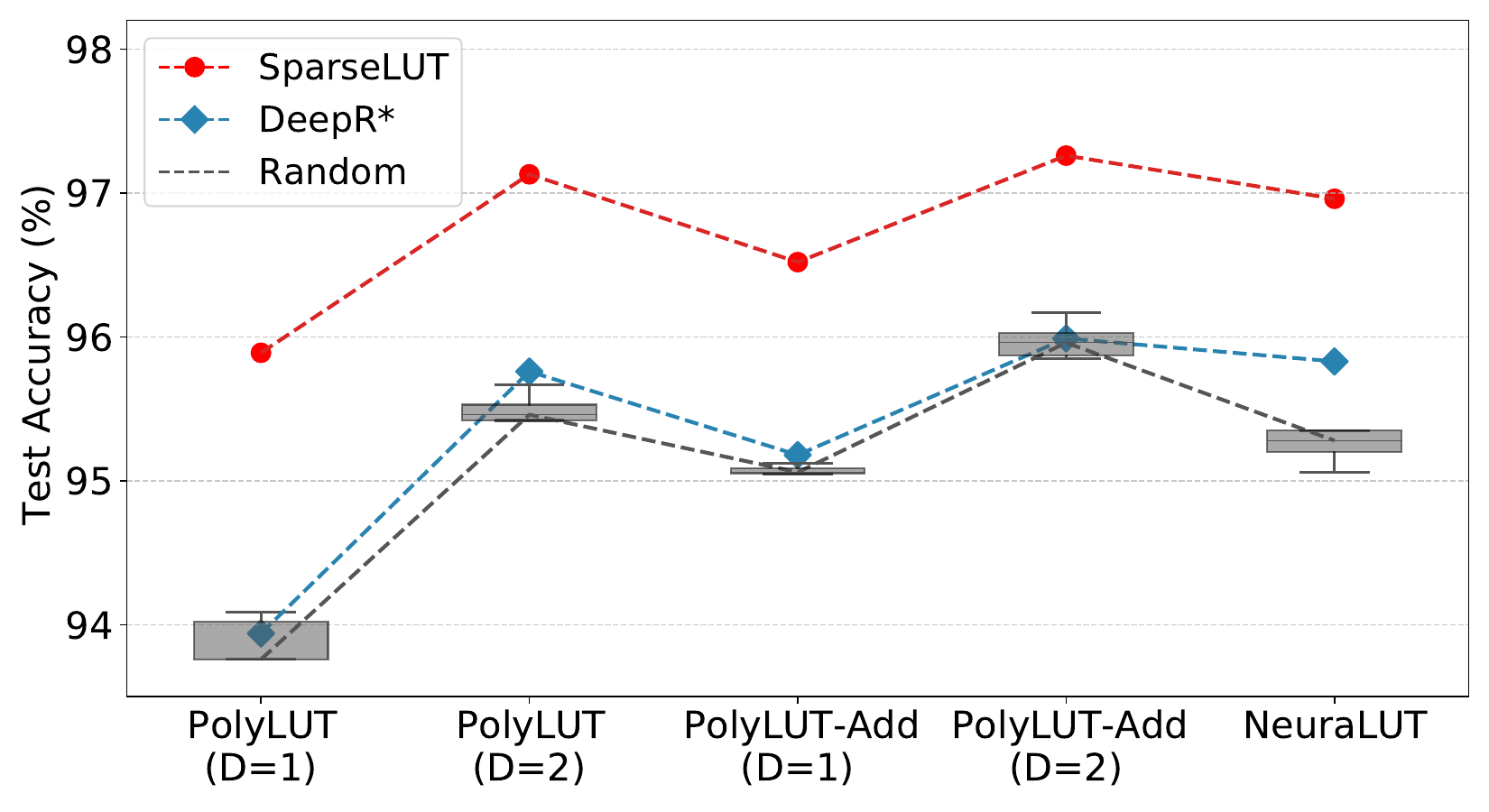}}
    \caption{Test accuracy for five models with random sparsity (5 seeds), DeepR$^*$, and SparseLUT. } 
    \label{fg:mnist_boxplot}
\end{figure}

Figure~\ref{fg:mnist_boxplot} further compares the test accuracy of applying different connectivity patterns to five models---PolyLUT ($D=1$), PolyLUT ($D=2$), PolyLUT-Add ($D=1$), PolyLUT-Add ($D=2$), and NeuraLUT---under three sparsity configurations: random sparsity (boxplots with five seeds), DeepR$^*$, and SparseLUT. For DeepR$^*$ and SparseLUT, the optimized accuracy points are plotted separately for each model.

The results show that both DeepR$^*$ and SparseLUT outperform random sparsity, but the advantages of SparseLUT are more pronounced. DeepR$^*$ achieves limited improvements over random sparsity in certain cases, such as PolyLUT ($D=1$) and PolyLUT-Add ($D=2$), where its accuracy points overlap with the box regions of random sparsity, despite being slightly above the mean line. In contrast, SparseLUT consistently delivers significant gains, exceeding the random sparsity range by 1.4\%--2.1\%. This performance gap highlights the effectiveness of SparseLUT’s non-greedy training approach, which provides a larger parameter search space in the early stages, allowing the model to explore a broader range of sparsity patterns.

\subsection{Evaluations}

\begin{table*}[]
  \centering
  \caption{Performance comparison with accuracy reported from prior works.}
  \label{tb:comparison}
  \bgroup
  \def\arraystretch{1.13}
  \setlength\tabcolsep{3.3mm}
  \scalebox{1.0}{
  \begin{tabular}{|c|l|l|c|l|c|c|}
  \cline{1-7}
  \multicolumn{1}{|c|}{\textbf{Dataset}}    & \multicolumn{1}{c|}{\textbf{Setup}} & \multicolumn{1}{c|}{\textbf{Model}} & \multicolumn{1}{c|}{\textbf{Accuracy}} & \multicolumn{1}{c|}{\textbf{Model}} & \multicolumn{1}{c|}{\textbf{Accuracy}} & \multicolumn{1}{c|}{\textbf{Dense}}\\ \hline \hline
  \multirow{5}{*}{MNIST}         &  HDR($D$=1)~\cite{polylut}                & PolyLUT      & 93.76\%   & SparseLUT-PolyLUT       & \bf{95.89\%}  & \multirow{4}{*}{98.55\%} \\
                                 &  HDR($D$=2)~\cite{polylut}                & PolyLUT      & 95.42\%   & SparseLUT-PolyLUT       & \bf{97.13\%}  & \\
                                 &  HDR-Add2($D$=1)~\cite{polylutadd}        & PolyLUT-Add  & 95.09\%   & SparseLUT-PolyLUT-Add   & \bf{96.52\%}  & \\
                                 &  HDR-Add2($D$=2)~\cite{polylutadd}        & PolyLUT-Add  & 95.87\%   & SparseLUT-PolyLUT-Add   & \bf{97.26\%}  & \\ \cline{7-7}
                                 &  HDR-5L~\cite{neuralut}                   & NeuraLUT     & 95.20\%   & SparseLUT-NeuraLUT      & \bf{96.96\%}  & 98.61\%\\ \hline\hline
  \multirow{5}{*}{JSC (high accuracy)} & JSC-XL($D$=1)~\cite{polylut}        & PolyLUT      & 74.48\%   & SparseLUT-PolyLUT       & \bf{74.65\%}  & \multirow{4}{*}{75.46\%} \\
                                 &  JSC-XL($D$=2)~\cite{polylut}             & PolyLUT      & 74.94\%   & SparseLUT-PolyLUT       & \bf{75.01\%}  &\\
                                 &  JSC-XL-Add2($D$=1)~\cite{polylutadd}     & PolyLUT-Add  & 74.64\%   & SparseLUT-PolyLUT-Add   & \bf{74.74\%}  &\\
                                 &  JSC-XL=Add2($D$=2)~\cite{polylutadd}     & PolyLUT-Add  & 74.98\%   & SparseLUT-PolyLUT-Add   & \bf{75.04\%}  &\\ \cline{7-7}
                                 &  JSC-5L~\cite{neuralut}                   & NeuraLUT     & 74.93\%   & SparseLUT-NeuraLUT      & \bf{74.98\%}  & 75.31\% \\ \hline\hline
   \multirow{5}{*}{JSC (low accuracy)} & JSC-M Lite($D$=1)~\cite{polylut}    & PolyLUT      & 71.65\%   & SparseLUT-PolyLUT       & \bf{72.10\%}  & \multirow{4}{*}{72.48\%}\\
                                 &  JSC-M Lite($D$=2)~\cite{polylut}         & PolyLUT      & 71.98\%   & SparseLUT-PolyLUT       & \bf{72.15\%}  &\\
                                 &  JSC-M Lite-Add2($D$=1)~\cite{polylutadd} & PolyLUT-Add  & 71.53\%   & SparseLUT-PolyLUT-Add   & \bf{71.96\%}  &\\
                                 &  JSC-M Lite-Add2($D$=2)~\cite{polylutadd} & PolyLUT-Add  & 71.90\%   & SparseLUT-PolyLUT-Add   & \bf{72.24\%}  &\\ \cline{7-7}
                                 &  JSC-2L~\cite{neuralut}                   & NeuraLUT     & 72.01\%   & SparseLUT-NeuraLUT      & \bf{72.95\%}  &73.34\%  \\ \hline

  \end{tabular}}
  \egroup

  \begin{tablenotes}
    \footnotesize
    \item[1] LogicNets~\cite{LogicNets} cases are equivalent to all baselines with $D=1$.
\end{tablenotes}
\end{table*}

In this subsection, we present the performance of SparseLUT across various baseline models. The accuracy values for random sparsity are sourced from referenced papers and open-source projects. As shown in Table~\ref{tb:comparison}, SparseLUT consistently outperforms random sparsity configurations across all baseline models.

To assess the ceiling for accuracy improvement achievable through optimized sparsity, we use the fully connected model of LogicNets ({\em i.e.}, polynomial degree $D=1$) as a fair comparison point for PolyLUT ($D=1$) and PolyLUT-Add ($D=1$) configurations. For models with $D>2$ or NeuraLUT's network-in-network architecture, fully connected implementations result in an exponential increase in parameters, rendering such implementations impractical. Consequently, the dense accuracy shown in the last column of Table~\ref{tb:comparison} for these configurations is considered a reasonable indicator rather than an upper bound.

The results highlight notable differences in accuracy deltas ($\delta$ = \textit{Dense Accuracy} - \textit{Random Sparsity Accuracy}) across datasets and models. For MNIST, SparseLUT achieves a substantial accuracy boost, attributed to the large $\delta$ exceeding 3\%. In contrast, for most models in the JSC dataset, this $\delta$ is much smaller—approximately 0.8\%—explaining the comparatively modest accuracy improvements achieved with optimized connectivity. A notable exception within the JSC dataset is the JSC-2L model on NeuraLUT (last row), which exhibits a $\delta$ of 1.33\%. SparseLUT significantly boosts its accuracy from 72.01\% to 72.95\%, likely due to the higher word length ($\beta = 4$) of the JSC-2L model compared to the 3-bit word length ($\beta = 3$) in other JSC-M Lite models.

These findings reveal that SparseLUT’s effectiveness is closely tied to the connectivity optimization potential of the underlying model, as reflected in the $\delta$ between dense and sparse configurations. This highlights the critical role of dataset characteristics and model architecture in determining the extent of accuracy improvements achievable with SparseLUT.

\begin{figure}[]
    \centerline{\includegraphics[width=1.0\linewidth]{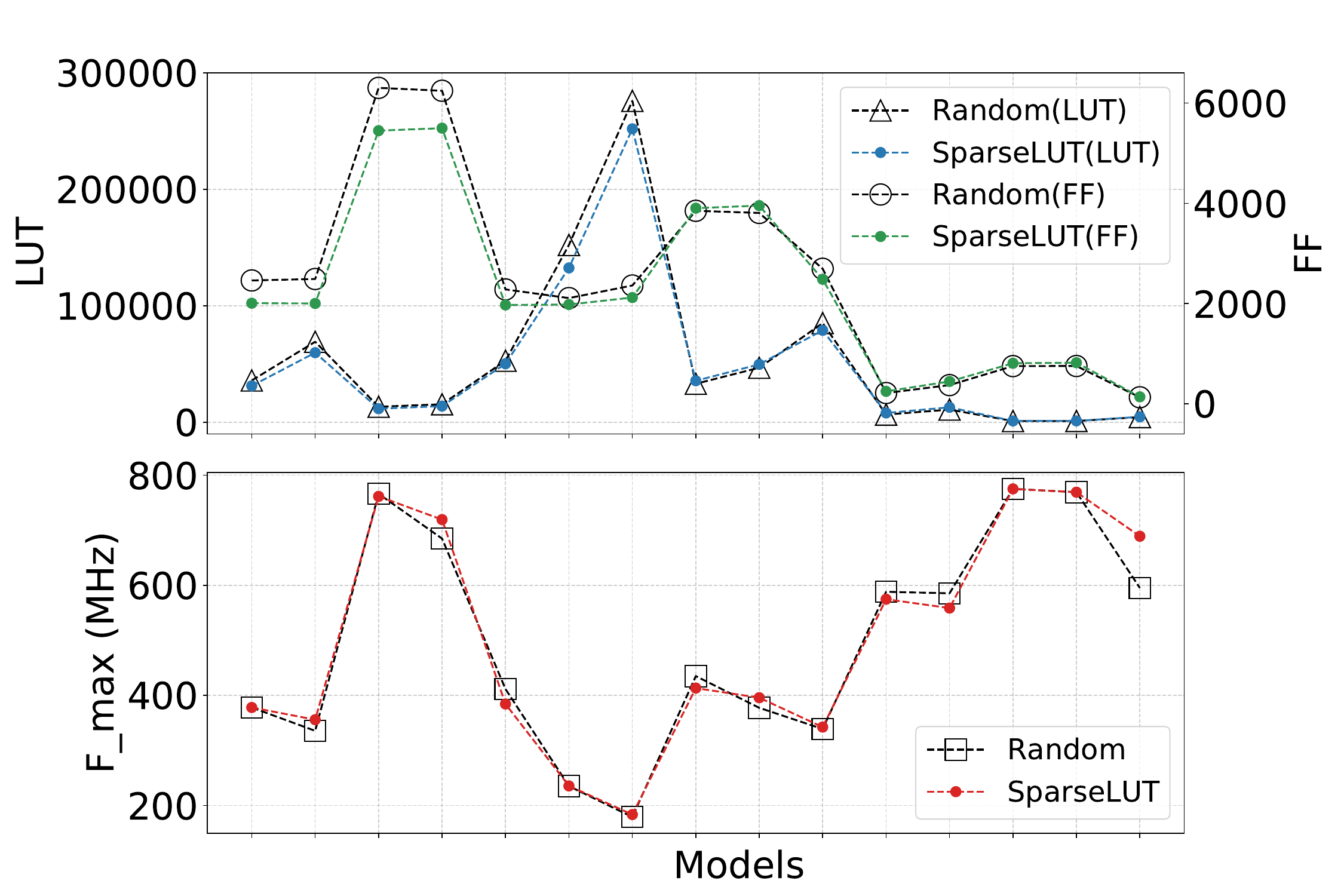}}
    \caption{Post place \& route hardware comparison between models (in the same order from left to right as Table~\ref{tb:comparison}) with/without connectivity from SparseLUT.} 
    \label{fg:lutff}
\end{figure}

SparseLUT does not alter the number of LUT entries in the generated RTL design; Figure~\ref {fg:lutff} illustrates the post-Place\&Route results for area (LUTs and flip-flops (FFs)) and maximum frequency (F\_max) from Vivado, constrained within the same clock cycle. Notably, SparseLUT achieves similar hardware consumption, and the F\_max comparison shows no speed penalty. Therefore, we conclude that the accuracy improvements achieved by SparseLUT are realized without incurring additional hardware and latency overhead.

\section{Conclusion}
\label{se:Conclusion}

This paper presented SparseLUT, a general dynamic sparsity training framework that improves the accuracy of LUT-based DNNs deployed on FPGAs. SparseLUT simply changes the connectivity between neurons in an FPGA implementation, and incurs no LUT or routing fabric overhead.

It was found that existing methods, such as LogicNets~\cite{LogicNets}, PolyLUT~\cite{polylutadd}, PolyLUT-Add~\cite{polylutadd}, and NeuraLUT~\cite{neuralut}, rely on random sparse connectivity patterns, limiting their accuracy potential. SparseLUT addresses this gap with a non-greedy training strategy that transitions from the progressive sparsification phase to fine-tuning while enforcing fixed fan-in constraints. Experimental results showed that SparseLUT outperformed all baseline methods, achieving up to a 2.13\% accuracy improvement on MNIST and 0.94\% on Jet Substructure Classification. We further demonstrated that SparseLUT consistently achieved better results than the state-of-the-art DeepR$^*$ technique by relaxing the requirement that the number of dropped and regrown connections must always match. Moreover, we used the dense LogicNets baseline as a trget for accuracy, providing a means to assess the benefits of optimized connectivity.

Future work could enhance SparseLUT by replacing its random regrowth mechanism with guided criteria, further improving accuracy. Integrating SparseLUT with advanced features such as the polynomial degree optimization of PolyLUT and the network-in-network structure of NeuraLUT during training could unlock additional potential for optimized sparsity.

\bibliographystyle{unsrt}
\bibliography{ref}

\end{document}